\documentclass[aps,prc,superscriptaddress,twoside,twocolumn,nofootinbib,10pt,%
showpacs, floatfix]{revtex4-1}

\usepackage{amsmath,amssymb}
\usepackage{graphicx,bm}
\usepackage{hyperref}
\usepackage{epstopdf}
\usepackage{ulem} 
\usepackage[usenames]{color}
\usepackage{float}

\allowdisplaybreaks

\newcommand{\non}{\nonumber}

\begin{document}

\title{\texorpdfstring{Heavy quark spin multiplet structure of $P_c(4312)$, $P_c(4440)$, and $P_c(4457)$}{Heavy quark spin multiplet structure of Pc(4312), Pc(4440), and Pc(4457)}}

\author{Yuki Shimizu}
\email{yshimizu@hken.phys.nagoya-u.ac.jp}
\affiliation{Department of Physics,  Nagoya University, Nagoya 464-8602, Japan}

\author{Yasuhiro Yamaguchi}
\email{yasuhiro.yamaguchi@riken.jp}
\affiliation{Theoretical Research Division, Nishina Center, RIKEN, Hirosawa, Wako, Saitama 351-0198, Japan}

\author{Masayasu Harada}
\email{harada@hken.phys.nagoya-u.ac.jp}
\affiliation{Department of Physics,  Nagoya University, Nagoya 464-8602, Japan}

\date{\today}


\begin{abstract}
Very recently, the LHCb collaboration has reported the new result about the hidden-charm pentaquarks:
$P_c(4312)$ near the $\bar{D}\Sigma_c$ threshold, and $P_c(4440)$ and $P_c(4457)$ near $\bar{D}^*\Sigma_c$ threshold. 
We study the heavy quark spin (HQS) multiplet structures of these newly $P_c$ pentaquarks 
under the heavy quark spin symmetry based on the hadronic molecular picture.
We point out that $P_c(4312)$ is the $J^P = 1/2^-$ member of an HQS triplet, and $P_c(4440)$ and $P_c(4457)$ are the $J^P = 3/2^-$ member of the HQS triplet and an HQS singlet with $J^P = 3/2^-$.
Namely, the $P_c(4312)$ and one of $P_c(4440)$ and $P_c(4457)$ belong to an HQS triplet.
The HQS multiplet structure predicts the existence of $J^P = 5/2^-$ state near $\bar{D}^\ast\Sigma_c^\ast$ threshold.

\end{abstract}

\maketitle


\section{Introduction}
\label{sec:Intro}
In 2015, the LHCb collaboration observed the two candidates of the hidden-charm pentaquark, $P_c(4380)$ and $P_c(4450)$
\cite{Aaij:2015tga, Aaij:2016phn, Aaij:2016ymb}.
Before the LHCb observation, some theoretical studies of hidden-charm pentaquarks were done~\cite{Wu:2010jy, Yang:2011wz, Wang:2011rga, Wu:2012md}.
The discovery of $P_c$ pentaquarks have further stimulated the study of the pentaquark.
In particular, the hadronic molecular interpretation 
\cite{Chen:2015loa, He:2015cea, Roca:2015dva, Meissner:2015mza, Shimizu:2016rrd, Yamaguchi:2016ote, He:2016pfa, Azizi:2016dhy}
is very famous because the masses of the $P_c$ pentaquarks are slightly below the $\bar{D}^{(*)}\Sigma_c^{(*)}$ thresholds.
There are also another explanation by the 
compact pentaquark \cite{Maiani:2015vwa, Lebed:2015tna, Li:2015gta, Wang:2015epa}, 
baryocharmonium model \cite{Kubarovsky:2015aaa}, 
quark-cluster model \cite{Takeuchi:2016ejt}, 
meson-baryon molecules coupled with five-quark states \cite{Yamaguchi:2017zmn}, 
and so on.

Very recently, the LHCb collaboration has reported the new result about the $P_c$ pentaquarks\cite{Skwarnicki:Moriond2019}.
There are three narrow peaks, $P_c(4312)$, $P_c(4440)$, and $P_c(4457)$.
Their masses and widths are 
\begin{align}
	P_c(4312) \hspace{15pt} M&=4311.9\pm0.7^{+6.8}_{-0.6}~{\rm MeV}, \non \\
	\Gamma&=9.8\pm2.7^{+3.7}_{-4.5}~{\rm MeV}, \non \\
	 P_c(4440) \hspace{15pt} M&=4440.3\pm1.3^{+4.1}_{-4.7}~{\rm MeV}, \non \\
	\Gamma&=20.6\pm4.9^{+8.7}_{-10.1}~{\rm MeV}, \non \\
	P_c(4457) \hspace{15pt} M&=4457.3\pm0.6^{+4.1}_{-1.7}~{\rm MeV}, \non \\
	\Gamma&=6.4\pm2.0^{+5.7}_{-1.9}~{\rm MeV}. \non
\end{align}
It is very interesting that a narrow peak of the $P_c(4450)$ in the previous result \cite{Aaij:2015tga, Aaij:2016phn, Aaij:2016ymb}
splits into two peaks ; $P_c(4440)$ and $P_c(4457)$.
Their masses are close to the $\bar{D}^*\Sigma_c$ threshold.
Moreover, there is a new narrow peak of $P_c(4312)$ near the $\bar{D}\Sigma_c$ threshold.
It is very natural to interpret that the $P_c(4312)$ is the $\bar{D}\Sigma_c$ molecule 
and the $P_c(4440)$ and $P_c(4457)$ are $\bar{D}^*\Sigma_c$ molecule.
Several discussions have already been made on these new pentaquark states
\cite{Chen:2019bip, Chen:2019asm, Guo:2019fdo, Liu:2019tjn, He:2019ify}.

The new result will be an important clue to understand the heavy quark spin symmetry (HQSS)
\cite{Isgur:1989vq, Isgur:1989ed, Isgur:1991wq, Neubert:1993mb, Manohar:2000dt}.
The degeneracy of the single heavy hadrons is well known.
For instance, the mass difference between $K$ and $K^*$ is about $400$~MeV.
On the other hand, the mass splitting between $D(B)$ and $D^*(B^*)$ is about $140(45)$~MeV.
The masses of the two hadrons with the different spin tend to degenerate as the quark mass increases.
The single heavy baryons like $\Sigma_c^{(*)}$ and $\Sigma_b^{(*)}$ have same tendency.
This is called the heavy quark spin (HQS) multiplet structure.
There are also the studies of the HQS multiplet structure of multi-hadron system with a heavy quark like heavy meson-nucleon molecular states\cite{Yasui:2013vca, Yamaguchi:2014era}.

In this paper, we discuss the HQS multiplet structure of new $P_c$ pentaquarks. 
The HQS multiplet of $\bar{P}^{(*)}\Sigma_Q^{(*)}$ molecular states has been studied in our previous works \cite{Shimizu:2018ran, Shimizu:2019jfy}.
The $\bar{P}$ and $\bar{P}^*$ denote heavy mesons of $J^P=0^-$ and $1^-$ with an anti-heavy quark, 
and the $\Sigma_Q$ and $\Sigma_Q^*$ denote the heavy baryons of $J^P=1/2^+$ and $3/2^+$ with a heavy quark.
We have defined the light-cloud spin (LCS) basis to deal with the HQSS for doubly heavy quark system and classified the HQS multiplet structure of the $P_c$-like pentaquarks.
We apply the analysis of the S-wave molecular states\cite{Shimizu:2018ran} to the new $P_c$ states.

This paper is organized as follows. 
In Sec.~\ref{sec:HQS multiplet}, we show the HQS multiplet structure of the S-wave $\bar{P}^{(*)}\Sigma_Q^{(*)}$ molecular states.
The theoretical interpretation for new $P_c$ states is given in Sec.~\ref{sec:Interpretation}.
Finally, we summarize this paper in Sec.~\ref{sec:summary}.

\section{\texorpdfstring{HQS multiplet structure of $\bar{P}^{(*)}\Sigma_Q^{(*)}$ molecular states}{HQS multiplet structure of Pbar(*) SigmaQ(*) molecular states}}
\label{sec:HQS multiplet}
In this section, we show the HQS multiplet of S-wave $\bar{P}^{(*)}\Sigma_Q^{(*)}$ molecular states  following  Ref.~\cite{Shimizu:2018ran}.
The possible spin states and meson-baryon components are shown in Table~\ref{tab:possible spin state}.
\begin{table}[!htbp]
\begin{center}
\caption{Possible spin states of
 the S-wave $\bar{P}^{(*)}\Sigma_Q^{(*)}$ molecular states with given $J^P$.
$\left(^{2S+1}L_{J}\right)$ denotes the total spin of meson and baryon $S$, the orbital angular momentum $L$, and the total angular momentum $J$.
}
\begin{tabular}{c|l}\hline
 $J^P$ &  \\ \hline
 $\frac{1}{2}^-$ & $\bar{P}\Sigma_{Q}\left( ^2S_{1/2} \right), \bar{P}^*\Sigma_{Q}\left( ^2S_{1/2} \right), \bar{P}^*\Sigma_{Q}^*\left( ^2S_{1/2} \right)$  \\[1mm] \hline
 $\frac{3}{2}^-$ & $\bar{P}\Sigma_{Q}^*\left( ^4S_{3/2} \right), \bar{P}^*\Sigma_{Q}\left( ^4S_{3/2} \right), \bar{P}^*\Sigma_{Q}^*\left( ^4S_{3/2} \right)$  \\[1mm] \hline
 $\frac{5}{2}^-$ & $\bar{P}^*\Sigma_{Q}^*\left( ^6S_{5/2} \right)$  \\[1mm] \hline
\end{tabular}
\label{tab:possible spin state}
\end{center}
\end{table}
Their spin structures in the hadronic molecule (HM) basis are given as
\begin{align}
    \bar{P}\Sigma_Q = 
    		\left[\bar{Q}q\right]_0 \otimes \left[Q[d]_1\right]_{1/2} &= \frac{1}{2}~,
    		\label{eq:spinPSigmaQ} \\
    \bar{P}\Sigma_Q^* = 
    		\left[\bar{Q}q\right]_0 \otimes \left[Q[d]_1\right]_{3/2} &= \frac{3}{2}~,
    		\label{eq:spinPSigmaQstar} \\
    \bar{P}^*\Sigma_Q = 
    		\left[\bar{Q}q\right]_1 \otimes \left[Q[d]_1\right]_{1/2} &= \frac{1}{2} \oplus \frac{3}{2}~,
    		\label{eq:spinPstarSigmaQ} \\
    \bar{P}^*\Sigma_Q^* = 
    		\left[\bar{Q}q\right]_1 \otimes \left[Q[d]_1\right]_{3/2} &= \frac{1}{2} \oplus \frac{3}{2} \oplus \frac{5}{2}~,
    		\label{eq:spinPstarSigmaQstar}
\end{align}
where the $\left[ \bar{Q}q \right]_{s_1}$ is a $\bar{P}^{(*)} \sim \bar{Q}q$ meson with spin $s_1$ 
and the $\left[ Q[d]_1 \right]_{s_2}$ is a $\Sigma_Q^{(*)}$ baryon with spin $s_2$.
The $[d]_1$ implies the light-diquark with spin $1$ in a $\Sigma_Q^{(*)}$.

Next, we define the LCS basis to deal with the HQSS.
The spin structures in the LCS basis are written as follows : 
\begin{align}
    \left[\bar{Q}Q\right]_0 \otimes \left[q[d]_1\right]_{1/2} &= \frac{1}{2}
     \hspace{5mm}({\rm singlet})~, \label{eq:spin structure 1/2 singlet} \\
    \left[\bar{Q}Q\right]_0 \otimes \left[q[d]_1\right]_{3/2} &= \frac{3}{2}
     \hspace{5mm}({\rm singlet})~, \label{eq:spin structure 3/2 singlet} \\
    \left[\bar{Q}Q\right]_1 \otimes \left[q[d]_1\right]_{1/2} &= \frac{1}{2} \oplus \frac{3}{2}
     \hspace{5mm}({\rm doublet})~, \label{eq:spin structure 1/2-3/2 doublet} \\
    \left[\bar{Q}Q\right]_1 \otimes \left[q[d]_1\right]_{3/2} &= \frac{1}{2} \oplus \frac{3}{2} \oplus \frac{5}{2}
      \hspace{5mm}({\rm triplet})~. \label{eq:spin structure 1/2-3/2-5/2 triplet}
\end{align}
There are four HQS multiplets for S-wave molecular states.
The basis transformation is done by 
\begin{align}
	\psi^{\rm LCS}_{J^P} = U^{-1}_{J^P}\psi^{\rm HM}_{J^P},
\end{align}
where the $U_{J^P}$ is a unitary matrix determined by the Clebsch-Gordan coefficients to rearrange the spin structure.
The wave functions and transformation matrices for each $J^P$ are given by as follows : 
\begin{align}
	\psi^{\textrm{HM}}_{1/2^-} &= \left(
	\begin{array}{l}
		\left| \bar{P}\Sigma_Q \right\rangle_{1/2^-} \\
		\left| \bar{P}^*\Sigma_Q \right\rangle_{1/2^-} \\
		\left| \bar{P}^*\Sigma_Q^* \right\rangle_{1/2^-}
	\end{array}
	\right)~, \label{eq:HMwave1/2} \\
	\psi^{\textrm{LCS}}_{1/2^-} &= U_{1/2^-}^{-1} \psi^{\textrm{HM}}_{1/2^-} \non \\
	 &= \left(
	\begin{array}{l}
		\left| \left[ \bar{Q}Q \right]_0 \otimes \left[ q[d]_1 \right]_{1/2} \right\rangle_{1/2^-}^{\textrm{singlet}} \\
		\left| \left[ \bar{Q}Q \right]_1 \otimes \left[ q[d]_1 \right]_{1/2} \right\rangle_{1/2^-}^{\textrm{doublet}} \\
		\left| \left[ \bar{Q}Q \right]_1 \otimes \left[ q[d]_1 \right]_{3/2} \right\rangle_{1/2^-}^{\textrm{triplet}}
	\end{array}
 \right)~, \label{eq:LCSwave1/2} \\
	U_{1/2^-} &= \left(
	\begin{array}{ccc}
		\frac{1}{2} & -\frac{1}{2\sqrt{3}} & \frac{2}{\sqrt{6}} \\
		-\frac{1}{2\sqrt{3}} & \frac{5}{6} & \frac{2}{3\sqrt{2}} \\
		\frac{2}{\sqrt{6}} & \frac{2}{3\sqrt{2}} & -\frac{1}{3}
	\end{array}
	\right)~, \label{eq:unitary matrix 1/2}
\end{align}
\begin{align}
	\psi^{\textrm{HM}}_{3/2^-} &= \left(
	\begin{array}{l}
		\left| \bar{P}\Sigma_Q^* \right\rangle_{3/2^-} \\
		\left| \bar{P}^*\Sigma_Q \right\rangle_{3/2^-} \\
		\left| \bar{P}^*\Sigma_Q^* \right\rangle_{3/2^-}
	\end{array}
	\right)~, \label{eq:HMwave3/2} \\
	\psi^{\textrm{LCS}}_{3/2^-} &= U_{3/2^-}^{-1} \psi^{\textrm{HM}}_{3/2^-} \non \\
	&= \left(
	\begin{array}{l}
		\left| \left[ \bar{Q}Q \right]_0 \otimes \left[ q[d]_1 \right]_{3/2} \right\rangle_{3/2^-}^{\textrm{singlet}} \\
		\left| \left[ \bar{Q}Q \right]_1 \otimes \left[ q[d]_1 \right]_{1/2} \right\rangle_{3/2^-}^{\textrm{doublet}} \\
		\left| \left[ \bar{Q}Q \right]_1 \otimes \left[ q[d]_1 \right]_{3/2} \right\rangle_{3/2^-}^{\textrm{triplet}}
	\end{array}
	\right)~, \label{eq:LCSwave3/2}  \\
	U_{3/2^-} &= \left(
	\begin{array}{ccc}
		\frac{1}{2} & -\frac{1}{\sqrt{3}} & \frac{\sqrt{15}}{6} \\
		-\frac{1}{\sqrt{3}} & \frac{1}{3} & \frac{\sqrt{5}}{3} \\
		\frac{\sqrt{15}}{6} & \frac{\sqrt{5}}{3} & \frac{1}{6}
	\end{array}
	\right)~, \label{eq:unitary matrix 3/2}
\end{align}
\begin{align}
	\psi^{\textrm{HM}}_{5/2^-} &= 
		\left| \bar{P}^*\Sigma_Q^* \right\rangle_{5/2^-}, \label{eq:HMwave5/2}\\
	\psi^{\textrm{LCS}}_{5/2^-} &= U_{5/2^-}^{-1} \psi^{\textrm{HM}}_{5/2^-} \non \\
	&= \left| \left[ \bar{Q}Q \right]_1 \otimes \left[ q[d]_1 \right]_{3/2} \right\rangle_{5/2^-}^{\textrm{triplet}}, \label{eq:LCSwave5/2} \\
	U_{5/2^-} &= 1~. \label{eq:unitary matrix 5/2}
\end{align}

In our previous study \cite{Shimizu:2018ran}, 
we constructed one-pion exchange potential (OPEP)  to determine which multiplet have an attractive potential.
As a result, the $J^P=3/2^-$ singlet of Eq.~\eqref{eq:spin structure 3/2 singlet} 
and the $J^P=(1/2^-, 3/2^-, 5/2^-)$ triplet of Eq.~\eqref{eq:spin structure 1/2-3/2-5/2 triplet} are attractive
when we assign $g=+0.59$.
We apply these HQS multiplet structures for the newly observed $P_c$ pentaquarks in the next section.

\section{\texorpdfstring{Theoretical interpretation for new $P_c$ states}{Theoretical interpretation for new Pc states}}
\label{sec:Interpretation}
In Sec.~\ref{sec:HQS multiplet}, we show the HQS multiplet states in the LCS basis.
These states can be written by the superposition of HM states.
For $J^P=1/2^-$,
\begin{align}
	| {\rm singlet} \rangle_{1/2^-} &= 
		\frac{1}{2} | \bar{P}\Sigma_Q \rangle_{1/2^-}
		- \frac{1}{2\sqrt{3}} | \bar{P}^*\Sigma_Q \rangle_{1/2^-} \non \\
		&\hspace{10pt}+ \frac{2}{\sqrt{6}} | \bar{P}^*\Sigma_Q^* \rangle_{1/2^-}, 
		\label{eq:singlet1/2 by superposition of HM} \\
	| {\rm doublet} \rangle_{1/2^-} &= 
		-\frac{1}{2\sqrt{3}} | \bar{P}\Sigma_Q \rangle_{1/2^-}
		+ \frac{5}{6} | \bar{P}^*\Sigma_Q \rangle_{1/2^-} \non \\
		&\hspace{10pt}+ \frac{2}{3\sqrt{2}} | \bar{P}^*\Sigma_Q^* \rangle_{1/2^-}, 
		\label{eq:doublet1/2 by superposition of HM} \\
	| {\rm triplet} \rangle_{1/2^-} &= 
		\frac{2}{\sqrt{6}} | \bar{P}\Sigma_Q \rangle_{1/2^-}
		- \frac{2}{3\sqrt{2}} | \bar{P}^*\Sigma_Q \rangle_{1/2^-} \non \\
		&\hspace{10pt}- \frac{1}{3} | \bar{P}^*\Sigma_Q^* \rangle_{1/2^-}. 
		\label{eq:triplet1/2 by superposition of HM}
\end{align}
For $J^P=3/2^-$,
\begin{align}
	| {\rm singlet} \rangle_{3/2^-} &= 
		\frac{1}{2} | \bar{P}\Sigma_Q^* \rangle_{3/2^-}
		- \frac{1}{\sqrt{3}} | \bar{P}^*\Sigma_Q \rangle_{3/2^-} \non \\
		&\hspace{10pt}+ \frac{\sqrt{15}}{6} | \bar{P}^*\Sigma_Q^* \rangle_{3/2^-}, 
		\label{eq:singlet3/2 by superposition of HM} \\
	| {\rm doublet} \rangle_{3/2^-} &= 
		-\frac{1}{\sqrt{3}} | \bar{P}\Sigma_Q^* \rangle_{3/2^-}
		+ \frac{1}{3} | \bar{P}^*\Sigma_Q \rangle_{3/2^-} \non \\
		&\hspace{10pt}+ \frac{\sqrt{5}}{3} | \bar{P}^*\Sigma_Q^* \rangle_{3/2^-}, 
		\label{eq:doublet3/2 by superposition of HM} \\
	| {\rm triplet} \rangle_{3/2^-} &= 
		\frac{\sqrt{15}}{6} | \bar{P}\Sigma_Q^* \rangle_{3/2^-}
		+ \frac{\sqrt{5}}{3} | \bar{P}^*\Sigma_Q \rangle_{3/2^-} \non \\
		&\hspace{10pt}+ \frac{1}{6} | \bar{P}^*\Sigma_Q^* \rangle_{3/2^-}. 
		\label{eq:triplet3/2 by superposition of HM}
\end{align}
For $J^P=5/2^-$,
\begin{align}
	| {\rm triplet} \rangle_{5/2^-} &= | \bar{P}^*\Sigma_Q^* \rangle_{5/2^-}. 
\end{align}
The component ratio is determined by the square of the each coefficient.
We summarize the  ratios 
in Table~\ref{tab:HM components for each LCS states}.
\begin{table}[!htbp]
\begin{center}
\caption{
Ratios of the hadronic molecular components for each HQS multiplet state.
Fifth column denotes whether the OPEP is attractive or repulsive.
}
\begin{tabular}{l|ccc|c}\hline
	$J^P=1/2^-$ & $\bar{P}\Sigma_Q$ & $\bar{P}^*\Sigma_Q$ & $\bar{P}^*\Sigma_Q^*$ & OPEP \\ \hline
	$|{\rm singlet}\rangle_{1/2^-}$ & 3 & 1 & 8 & repulsive \\
	$|{\rm doublet}\rangle_{1/2^-}$ & 3 & 25 & 8 & repulsive \\
	$|{\rm triplet}\rangle_{1/2^-}$ & 6 & 2 & 1 & attractive \\ \hline
\end{tabular}
\begin{tabular}{l|ccc|c}\hline
	$J^P=3/2^-$ & $\bar{P}\Sigma_Q^*$ & $\bar{P}^*\Sigma_Q$ & $\bar{P}^*\Sigma_Q^*$ & OPEP \\ \hline
	$|{\rm singlet}\rangle_{3/2^-}$ & 3 & 4 & 5 & attractive \\
	$|{\rm doublet}\rangle_{3/2^-}$ & 3 & 1 & 5 & repulsive \\
	$|{\rm triplet}\rangle_{3/2^-}$ & 15 & 20 & 1 & attractive \\ \hline
\end{tabular}
\begin{tabular}{l|c|c}\hline
	$J^P=5/2^-$ & $\bar{P}^*\Sigma_Q^*$ & OPEP \\ \hline
	$|{\rm triplet}\rangle_{5/2^-}$ & 1 & attractive \\ \hline
\end{tabular}
\label{tab:HM components for each LCS states}
\end{center}
\end{table}

Focusing on the $|{\rm triplet}\rangle_{1/2^-}$ state, the main component is the $\bar{P}\Sigma_Q$ molecular state.
In charm region, the threshold value of the $\bar{D}\Sigma_c$ is about $4320$~MeV.
It is natural to consider that $|{\rm triplet}\rangle_{1/2^-}$ corresponds to the $P_c(4312)$.
In fact, we have obtained a bound state of $|{\rm triplet}\rangle_{1/2^-}$ in our previous study \cite{Shimizu:2018ran}.

For $J^P=3/2^-$, the $|{\rm singlet}\rangle_{3/2^-}$ and $|{\rm triplet}\rangle_{3/2^-}$ are
the admixture of three and two states, respectively.
We did not get any solution around $\bar{D}^*\Sigma_c$ threshold \cite{Shimizu:2018ran}
because we calculated only bound state solutions.
It is needed to search poles of resonance states in the complex energy plane using e.g., 
the complex scaling method.
We expect that two Feshbach-like resonance states, $|{\rm singlet}\rangle_{3/2^-}$ and $|{\rm triplet}\rangle_{3/2^-}$, exist around $\bar{D}^*\Sigma_c$ threshold, 
and the one is the $P_c(4440)$ and the other is the $P_c(4457)$.

For $J^P=5/2^-$, only one state exists around $\bar{D}^*\Sigma_c^*$ for S-wave molecular state.
This is very interesting state as a third partner of the HQS triplet.
However, D-wave decay is necessary because S-wave $J/\psi p$ can not couple to the $5/2^-$ state.

\section{Summary and discussion}
\label{sec:summary}
In this work, we  studied the HQS multiplet structure of 
$P_c(4312)$, $P_c(4440)$ and $P_c(4457)$, which are newly reported from 
the LHCb collaboration.
We focused the $J^P=3/2^-$ singlet and $J^P=(1/2^-, 3/2^-, 5/2^-)$ triplet which have the attractive potential under the OPEP.

The $J^P=1/2^-$ is attractive only in the state belonging to the HQS triplet.
The most natural interpretation is that this state is corresponding to the $P_c(4312)$.

The HQS singlet and triplet are attractive for $J^P=3/2^-$.
The former is a mixture of the $\bar{D}\Sigma_c^*$, $\bar{D}^*\Sigma_c$ and $\bar{D}^*\Sigma_c^*$,
and the latter is a mixture of mainly the $\bar{D}\Sigma_c^*$ and $\bar{D}^*\Sigma_c$.
It is considered that the $P_c(4440)$ and $P_c(4457)$ can be obtained as the resonance state of the two multiplets.
In our previous work\cite{Shimizu:2018ran}, 
$|{\rm singlet}\rangle_{3/2^-}$ and $|{\rm triplet}\rangle_{3/2^-}$ are degenerate at the heavy quark limit.
However, $|{\rm singlet}\rangle_{3/2^-}$ state is slightly heavier at the finite quark mass by the breaking effect of the HQSS.
Therefore, the $P_c(4457)$ may be the HQS singlet state.
In order to investigate these states, 
it is necessary to investigate poles on the complex energy plane precisely.

The $J^P=5/2^-$ state of the HQS triplet is very interesting as a strong evidence to verify the HQS multiplet of hadronic molecule.
We expect that there exist a $\bar{D}^*\Sigma_c^*$ bound state below its threshold.

The spin and parity are not determined in the LHCb report.
Then, the other $J^P$ assignment can not be excluded.
The information of the spin and parity of $P_c$ pentaquarks is very useful clue to understand the HQS multiplet structure of hadronic molecular states.
At the same time, searching the missing HQS partners is also an important task.
We expect to see more details in the future experiments.

\acknowledgments
The work of Y.S. is supported in part by JSPS Grant-in-Aid for JSPS Research Fellow No. JP17J06300. 
The work of M.H. is supported in part by 
JPSP KAKENHI
Grant Number 16K05345. 
The work of Y.Y. is supported in part by the Special Postdoctoral Researcher (SPDR) and iTHEMS Programs of RIKEN.

\end{document}